\documentclass[preprint,10pt]{elsarticle}
\biboptions{numbers,sort&compress}
\usepackage{amsmath,amsfonts}
\usepackage{algorithmic}
\usepackage{algorithm}
\usepackage{array}
\usepackage{textcomp}
\usepackage{stfloats}
\usepackage{url}
\usepackage{verbatim}
\usepackage{graphicx}
\usepackage{dcolumn}
\usepackage{amsmath}
\usepackage{amsfonts}
\usepackage{mathrsfs}
\usepackage{physics}
\usepackage{bm}
\usepackage{dsfont}
\usepackage{enumerate}
\usepackage{slashed}
\usepackage[dvipsnames]{xcolor}
\usepackage{pgfplots}
\pgfplotsset{compat=1.16}

\usepackage{epsfig}
\usepackage{bm}
\usepackage{amssymb}
\usepackage{amsmath}
\usepackage{soul}

\usepackage{lipsum}
\usepackage{bbm}
\usepackage{amsthm}

\usepackage{xcolor}
\usepackage[colorlinks=true, citecolor=blue, linkcolor=red, urlcolor=magenta]{hyperref}
\newcommand{\one}{\mathbbm{1}}

\begin{document}

\begin{frontmatter}

\title{Polarization, Maximal Concurrence, and Pure States in High-Energy Collisions}

\author[first]{Yu-Xuan Liu}
\ead{121090366@link.cuhk.edu.cn}

\author[second]{Wei Qi}
\ead{qi.673@osu.edu}
\author[first]{Luo-Ting He}
\author[first,third]{Bo-Wen Xiao}
\ead{xiaobowen@cuhk.edu.cn}
\address[first]{School of Science and Engineering, The Chinese University of Hong Kong (Shenzhen), Longgang, Shenzhen, 518172, Guangdong, P.R. China}

\address[second]{Department of Physics, The Ohio State University, Columbus, OH 43210, USA}

\address[third]{Southern Center for Nuclear-Science Theory (SCNT), Institute of Modern Physics, Chinese Academy of Sciences, Huizhou, 516000, Guangdong, P.R. China}

\begin{abstract}
We establish a quantitative relation between local spin polarization and 
quantum entanglement in two-qubit systems by mathematically proving a closed-form upper bound on 
the concurrence at fixed local polarization magnitudes. The bound shows how the maximal concurrence is jointly constrained by the two local polarizations. We further 
demonstrate that this bound is saturated by pure states in certain cases with identical polarizations. As a concrete physical application, we consider the parity-violating 
process $e^+e^- \to Z^0 \to q\bar{q}$, which generates final-state spin 
polarization. We show that the maximal concurrence is attained in specific 
kinematic regions and is significantly reduced relative to the unpolarized 
case. These results establish a general, process-independent framework 
connecting local polarization, maximal entanglement, and pure states.
\end{abstract}


\end{frontmatter}

%\maketitle

\section{Introduction}
\label{introduction}

The spin degree of freedom of quarks and antiquarks produced in high-energy collisions
encodes rich information about the underlying dynamics of both perturbative and
nonperturbative QCD, as well as electroweak interactions. Understanding how polarization
arises in quark-antiquark pair production is not only central to the physics of spin
in QCD, but also provides a natural arena for exploring quantum correlations, especially the entanglement between the spin states of the produced pair. In this work, we explore the relation that connects the polarization of the
quark-antiquark pair to the maximal concurrence, a measure of spin entanglement,
and apply it to several physically relevant production mechanisms. Quark polarization in the final state can originate from a variety of sources,
which we briefly survey below to introduce the generality of this study. 
\begin{enumerate}
    \item \textbf{Polarized deep inelastic scattering:} When a polarized lepton scatters off a polarized nucleon, the struck quark inherits
a longitudinal polarization governed by the helicity parton distribution functions
$\Delta q(x, Q^2)$. The polarization transfer from the initial-state nucleon to the
final-state quark has been extensively studied in polarized deep inelastic scattering
(DIS) experiments, providing the primary means of extracting the spin structure of
the nucleon~\cite{EuropeanMuon:1987isl,HERMES:2006jyl}.

\item \textbf{Parity-violating processes:} The chiral nature of the electroweak interaction provides a clean mechanism for producing polarized quarks even from unpolarized initial states. In $e^+e^-$
annihilation through a $Z^0$ boson~\cite{ALEPH:2005ab}, the produced quark-antiquark pair acquires
a net polarization determined by the weak mixing angle and the
quark electroweak quantum numbers.

\item \textbf{Spin-orbit correlations:}
In unpolarized hadron collisions, where neither initial-state polarization nor parity violation is present, the fragmentation process itself can generate polarizations through final-state interactions that couple spin to the production-plane geometry, as evidenced by the large transverse polarization of $\Lambda$ hyperons observed in unpolarized $pp$ collisions~\cite{Bunce:1976yb}. The underlying mechanism for the self-polarization of $\Lambda$ hyperons still remains a long-standing and interesting question in QCD.

\item \textbf{Global polarization in heavy-ion collisions:}
In non-central heavy-ion collisions, the large orbital angular momentum of the
colliding system generates a vortical quark-gluon plasma. Through spin-orbit
coupling in the medium, quarks and antiquarks acquire a net polarization aligned
with the system's angular momentum~\cite{Liang:2004ph}. This \emph{global polarization} has been
observed via the polarization of $\Lambda$ and $\bar{\Lambda}$ hyperons by the
STAR Collaboration at RHIC~\cite{STAR:2017ckg, STAR:2018gyt}, establishing the
quark-gluon plasma as the most vortical fluid ever observed.
\end{enumerate}

The different mechanisms summarized above demonstrate that polarized quark-antiquark
pairs are produced across a wide range of collision systems and energy scales.
In recent years, a new perspective has emerged: the spin state of the $q\bar{q}$
pair can be analyzed using the tools of quantum information theory, with
entanglement serving as a quantitative probe of the quantum correlations generated
by the production dynamics~\cite{Uwer:2004vp, Hentschinski:2023izh, Afik:2020onf, Afik:2022dgh, Afik:2022kwm, Cheng:2024btk, Wu:2024mtj, Barr:2024djo, Ehataht:2023zzt, Guo:2024jch, Morales:2023gow, Parke:1996pr, Altakach:2022ywa, Aoude:2024xpx, Bernal:2023jba, Brandenburg:2002xr,  Hentschinski:2024gaa, James:2001klt, Kowalska:2024kbs, Li:2023gkh, Maltoni:2024csn, Severi:2022qjy, Subba:2024mnl, Severi:2021cnj, Wu:2024asu, Li:2024luk, Mahlon:1995zn, Dong:2023xiw, Aguilar-Saavedra:2022uye, Fabbrichesi:2023cev, Ghosh:2023rpj, White:2024nuc, Han:2023fci, Du:2024sly, Morales:2024jhj, Yang:2024kjn, Fabbrichesi:2021npl, Aoude:2022imd, Fabbrichesi:2024wcd, Afik:2025ejh, Fabbrichesi:2025psr,Cheng:2025cuv, Qi:2025onf, Cheng:2025zaw, Lin:2025eci, Low:2025aqq, Hatta:2025obw,Xi:2025feb, Gu:2025ijz, Guo:2026yhz,Fucilla:2025kit, Zhang:2025ean, Zhang:2026nwm,Fucilla:2026mkg}. Remarkably, ATLAS~\cite{ATLAS:2023fsd} and CMS~\cite{CMS:2024pts} collaborations have successfully measured quantum entanglement between top quark pairs in proton-proton collisions at the Large Hadron Collider. Recently, STAR collaboration~\cite{STAR:2025njp} measured the quantum spin entanglement between $\Lambda \bar{\Lambda}$ at the Relativistic Heavy-Ion Collider, and opened a new experimental window for studying decoherence and QCD effects.

The concurrence~\cite{Wootters:1997id, Hill:1997pfa} provides a particularly
convenient entanglement measure for bipartite spin-$\tfrac{1}{2}$ systems.
As we intend to show, the maximal value of the concurrence, referred to as the \emph{maximal concurrence}, captures the intrinsic entanglement content of the pair and is governed by the polarizations of the quark and antiquark in the final state. More specifically, as the polarizations increase, more spin information is stored locally in each quark, thereby constraining the amount of nonlocal quantum correlation that can be shared by the $q\bar{q}$ pair, and consequently the maximal concurrence is constrained by the two local polarization magnitudes. Furthermore, we find that maximizing the concurrence subject to certain physical constraints yields a density matrix corresponding to a pure state, revealing a deep connection between polarization, purity, and entanglement.

This paper is devoted to deriving and analyzing this relation. Our formalism
applies to any $q\bar{q}$ production mechanism and provides a unified
theoretical framework for understanding how single-particle polarization
and pairwise maximal entanglement are interrelated. This interplay is governed by
the mathematical structure of the density matrix and the underlying
interaction.

The connection between polarization and quantum entanglement has attracted
growing attention. In the context of polarized deep inelastic scattering,
Ref.~\cite{Cheng:2025zaw} recently explored quantum information observables
at the Electron-Ion Collider with transverse beam polarization, demonstrating
that initial-state spin configurations leave distinct imprints on the
entanglement structure of the final state. For parity-violating processes,
Ref.~\cite{Guo:2026yhz} performed a comprehensive quantum tomography of
fermion pairs in $e^+e^-$ collisions. Our study is complementary to these
efforts: we establish a general, process-independent relation between the
polarization of the individual quark and antiquark and the maximal concurrence
of the pair, and then illustrate the results with an explicit calculation
for the physical process $e^+e^- \to Z^0 \to q\bar{q}$, identifying the kinematic
regimes in which maximal entanglement is achieved. Looking ahead, extending
such analyses to incorporate transverse spin asymmetries in semi-inclusive
processes and global polarization measurements in heavy-ion collisions
may open new windows onto nonperturbative QCD dynamics through the
lens of quantum information.

\section{Formalism for two-Qubit system}
A general two-qubit system can be represented as a $4 \times 4$ density matrix, written as
\begin{equation}
\rho = \frac{1}{4} \left(
\one_{4}
+  B_i^{+} \, \sigma_i \otimes \one_{2}
+  B_j^{-} \, \one_{2} \otimes \sigma_j
+  C_{ij} \, \sigma_i \otimes \sigma_j
\right),
\label{density}
\end{equation}
where \textbf{$B_i^{+} = \operatorname{Tr}\!\left[\rho(\sigma_i \otimes \one_2)\right]$} and \textbf{$B_j^{-} = \operatorname{Tr}\!\left[\rho(\one_2 \otimes \sigma_j)\right]$} form the Bloch vectors of the two qubits and represent the individual spin polarizations, while \textbf{$C_{ij} = \operatorname{Tr}\!\left[\rho(\sigma_i \otimes \sigma_j)\right]$} is the correlation matrix characterizing the spin correlations between the qubits.

The degree of mixedness of a quantum state is quantified by its purity, $\mathrm{Tr}(\rho^2)$.
For a pure state, $\mathrm{Tr}(\rho^2)=1$, whereas for the maximally mixed state 
$\rho = \tfrac{1}{4}\one_{4\times4}$, the purity is $\tfrac{1}{4}$.

Concurrence $\mathcal{C}$ provides a quantitative measure of the entanglement. For any two-qubit system, the concurrence is given by~\cite{Wootters:1997id}:
\begin{equation}
\mathcal{C}[\rho]
= \max\{0,r_1-r_2-r_3-r_4\},
\end{equation}
where ${r_i}$ are the \textit{square roots} of the eigenvalues of the matrix $\mathcal R =\rho\tilde{\rho}$, ordered as $r_1 \ge r_2 \ge r_3 \ge r_4$. The concurrence ranges from 0 to 1, with larger values indicating a higher degree of entanglement. Here $\tilde{\rho}$ is the spin-flipped state of the density matrix $\rho$, defined as
\begin{equation}
\tilde{\rho}
= (\sigma_y \otimes \sigma_y)\,\rho^{\ast}\,(\sigma_y \otimes \sigma_y).
\end{equation}
For pure two-qubit states, the concurrence can be equivalently expressed in a simpler way, which is the so-called I-concurrence~\cite{Rungta:2001zcj} defined as:
\begin{equation}
\label{eq:I}
\mathcal{C}[\rho] =  \mathcal{C}_I[\rho]
= \sqrt{\,2\bigl(1-\operatorname{Tr}\rho_A^{\,2}\bigr)}
= \sqrt{\,2\bigl(1-\operatorname{Tr}\rho_B^{\,2}\bigr)}.
\end{equation}
where $\rho_A =\operatorname{Tr}_B(\rho)$ and $\rho_B=\operatorname{Tr}_A(\rho)$ are the reduced density matrices.

An important and widely used subclass of two-qubit states is the X state, whose density matrix has nonzero entries only on the main diagonal and anti-diagonal:
\begin{equation}
\rho_X=
\begin{pmatrix}
\rho_{11} & 0 & 0 & \rho_{14} \\
0 & \rho_{22} & \rho_{23} & 0 \\
0 & \rho_{23}^{\ast} & \rho_{33} & 0 \\
\rho_{14}^{\ast} & 0 & 0 & \rho_{44}
\end{pmatrix}.
\label{eq:Xstate}
\end{equation}
For $\rho_X$ to be a physical density matrix, it must satisfy the following conditions:
\begin{equation}
\rho_{ii}\in\mathbb{R},\quad \rho_{ii}\ge 0\ (i=1,2,3,4),\quad
\sum_{i=1}^{4}\rho_{ii}=1,
\end{equation}
and
\begin{equation}
|\rho_{14}|^2\le \rho_{11}\rho_{44},
\qquad
|\rho_{23}|^2\le \rho_{22}\rho_{33},
\label{eq:Xstate-positivity}
\end{equation}
which ensure the positivity of the density matrix~\cite{Ali:2010unc}. For this class, the concurrence takes the compact form~\cite{Wang:2006fru}
\begin{equation}
\mathcal{C}[\rho_X]=2\max\!\left\{0,\,|\rho_{14}|-\sqrt{\rho_{22}\rho_{33}},\,|\rho_{23}|-\sqrt{\rho_{11}\rho_{44}}\right\}.
\label{eq:Cxstate}
\end{equation}

\section{Maximal concurrence of $X$ states}
\label{secpolar}
In a two-qubit system, polarization can affect the entanglement and limit the maximal concurrence attainable by the system. Here we derive an analytic relation between the magnitudes of the two local polarizations $(\|\vec{B}^{\,+}\|$ and $\|\vec{B}^{\,-}\|)$, and the maximal concurrence. For convenience, define
\begin{equation}
        \mathbf a=\vec{B}^{\,+},\qquad \mathbf b=\vec{B}^{\,-},
        \qquad
        a=\|\mathbf a\|,\qquad b=\|\mathbf b\|,
\end{equation}

Since local unitary transformations preserve the concurrence and the
magnitudes of the local Bloch vectors, we may choose local bases in
which these vectors point along the positive \textit{z}-axis. This basis
choice does not, however, imply that a general two-qubit density matrix
has X form. Conversely, for the X-states introduced in Eq.~(\ref{eq:Xstate}), the
reduced density matrices are already diagonal in the computational
basis, so their Bloch vectors are already aligned with the
\textit{z}-axis; local basis flips can be used, if necessary, to make
their \textit{z}-components nonnegative. Therefore, without loss of generality, we can use local unitary transformations to rotate the Bloch vectors to the $z$-direction, i.e., $\mathbf{a}= (0,0,a)$ and $\mathbf{b}=(0,0,b)$.

We first optimize the concurrence within the X-state family to obtain
explicit states attaining the proposed bound. This is an additional
restriction, distinct from the choice of local bases, and does not by
itself establish the bound for arbitrary states. In Sec.~4, we prove
that the same expression is an upper bound for all two-qubit density
matrices, thereby establishing its global optimality at fixed local
polarization magnitudes.

%We focus on the magnitudes of the polarizations rather than their directions, since local unitary transformations preserve both the concurrence and the magnitudes of the local polarizations. Therefore, without loss of generality, we can use local unitary transformations to rotate the Bloch vectors to the $z$-direction, i.e., $\mathbf{a}= (0,0,a)$ and $\mathbf{b}=(0,0,b)$.
%We write:
%\begin{equation}
%\vec{B}^{\,+}=(0,0,a),\qquad \vec{B}^{\,-}=(0,0,b).
%\end{equation}
%where $a=\|\vec{B}^{\,+}\|$, $b=\|\vec{B}^{\,-}\|$, and $0\le a,b\le 1$. 

To build intuition, we first analyze the optimization within the class of
$X$ states, where the concurrence and positivity constraints take a simple
form. This allows us to construct explicit states that saturate the bound.
In the next section, we give a mathematical proof that the same closed-form
expression is an upper bound for arbitrary two-qubit density matrices.

For the $X$ state in Eq.~\eqref{eq:Xstate}, the local polarizations satisfy
\begin{equation}
a=(\rho_{11}+\rho_{22})-(\rho_{33}+\rho_{44}),
\quad
b=(\rho_{11}+\rho_{33})-(\rho_{22}+\rho_{44}).
\end{equation}
Since one degree of freedom remains in the diagonal entries after fixing $a$ and $b$, we introduce an additional real parameter $c$ to parametrize them:
\begin{equation}
c=\rho_{11}-\rho_{22}-\rho_{33}+\rho_{44}.
\end{equation}
Then the diagonal entries can be written as
\begin{align}
\rho_{11}&=\frac{1+a+b+c}{4},
&\rho_{22}&=\frac{1+a-b-c}{4},\\
\rho_{33}&=\frac{1-a+b-c}{4},
&\rho_{44}&=\frac{1-a-b+c}{4}.
\end{align}
Combining Eqs.~\eqref{eq:Xstate-positivity} and \eqref{eq:Cxstate}, we obtain $\mathcal{C}^{(X)} \leq 2\left|\sqrt{\rho_{11}\rho_{44}}-\sqrt{\rho_{22}\rho_{33}}\right|$. The upper bound is saturated when $\rho_{14}$ and $\rho_{23}$ saturate the positivity constraints, and the maximal concurrence can then be written as a function of $c$:
\begin{equation}
\mathcal{C}_{\max}^{(X)}(c)=2\left|\sqrt{\rho_{11}\rho_{44}}-\sqrt{\rho_{22}\rho_{33}}\right|.
\end{equation}
Using the above parameterization, one finds
\begin{align}
\rho_{11}\rho_{44}
=\frac{(1+c)^2-(a+b)^2}{16}, \quad \quad 
\rho_{22}\rho_{33}
=\frac{(1-c)^2-(a-b)^2}{16},
\end{align}
and therefore
\begin{equation}
\mathcal{C}_{\max}^{(X)}(c)
=
\frac12
\left|
\sqrt{(1+c)^2-(a+b)^2}
-
\sqrt{(1-c)^2-(a-b)^2}
\right|.
\end{equation}

From the diagonal positivity condition, the allowed interval is
\begin{equation}
a+b-1 \le c \le 1-|a-b|.
\label{eq:range_c}
\end{equation}
Since $\sqrt{\rho_{11}\rho_{44}}$ increases with $c$ and $\sqrt{\rho_{22}\rho_{33}}$ decreases with $c$, the maximum is attained at one of the endpoints in Eq.~\eqref{eq:range_c}. After comparing the values of $\mathcal{C}$ at the two endpoints, we find that the maximum with fixed local polarizations is achieved at $c=1-|a-b|$, with the value
\begin{equation}
\mathcal{C}_{\max}^{(X)}(a,b)
=
\sqrt{(1-\max\{a,b\})(1+\min\{a,b\})}.
\label{eq:CmaxXab}
\end{equation}
The corresponding optimal density matrices are listed in \ref{app:optimal-x} and are consistent with the optimal density matrix for fixed reduced density matrices (marginals) obtained in Ref.~\cite{Baio:2019cgb}. 

\section{Maximal concurrence of general states}
Next, we show that this new closed-form upper bound for $X$ states also holds for general states.
\begin{proof}
Let the eigenvalues of $\rho$ be $\lambda_1\ge\lambda_2\ge\lambda_3\ge\lambda_4\ge 0$.
The reduced states are
\begin{equation}
       \rho_A = \Tr_B \rho = \frac{1}{2}(\one_2 + \mathbf a \cdot \boldsymbol \sigma), \quad
       \rho_B = \Tr_A \rho = \frac{1}{2}(\one_2 + \mathbf b \cdot \boldsymbol \sigma).
\end{equation}
Note that the directions of $\mathbf{a}$ and $\mathbf{b}$ can be arbitrary and the conclusions that follow are not restricted to $X$ states.
Their lowest eigenvalues are
\begin{equation}
        \lambda_A = \frac{1-a}{2}, \quad \lambda_B = \frac{1-b}{2}.
\end{equation}

By the result in~\cite[Eq.~(2)]{Bravyi:2003vny},
\begin{equation}
       |\lambda_A-\lambda_B| \le \min\left( \lambda_1-\lambda_3, \lambda_2-\lambda_4 \right),
\end{equation}
which immediately implies
\begin{equation}
       \lambda_2 \ge \lambda_2-\lambda_4 \ge  | \lambda_A-\lambda_B | 
       = \frac{|a-b|}{2}.
\end{equation}
To obtain the desired upper bound on the concurrence, we maximize $\Tr(\rho^2) = \lambda_1^2+\lambda_2^2+\lambda_3^2+\lambda_4^2$
subject to the constraints
\begin{align}
        \lambda_1+\lambda_2+\lambda_3+\lambda_4 &= 1, \\ 
        \lambda_1 \ge \lambda_2 &\ge \frac{|a-b|}{2}, \\
        \lambda_3 \ge \lambda_4 &\ge 0.
\end{align}
The maximum is reached at
\begin{equation}
        \lambda_1 = 1- \frac{|a-b|}{2}, 
        \quad 
        \lambda_2 = \frac{|a-b|}{2}, 
        \quad 
        \lambda_3 = \lambda_4 = 0.
        \label{eq:eigenvaluesOfRho}
\end{equation}
Substitution gives
\begin{align}
        \max \Tr(\rho^2) 
        &= \left( 1-\frac{|a-b|}{2} \right)^2 
        + \left( \frac{|a-b|}{2} \right)^2 \notag\\
        &= 1 - |a-b| - ab 
        +\frac{1}{2}\left(a^2+b^2\right).
\end{align}
It follows that
\begin{align}
       \max \Tr(\rho \tilde \rho) 
       &= \max \Tr(\rho^2)-\frac{1}{2}\left( a^2+b^2 \right) \notag\\
       &= 1 - |a-b|-ab.
\end{align}
Therefore,
\begin{equation}
        \mathcal C^2[\rho] \le \Tr(\rho \tilde \rho) 
        \le 1-|a-b|-ab .
        \label{eq:C2bound}
\end{equation}
Equivalently, this can be written as
\begin{equation}
   \mathcal C[\rho] \le \sqrt{(1-\max\{a,b\})(1+\min\{a,b\})},
   \label{eq:CboundAppendix}
\end{equation}
which gives the bound of Eq.~\eqref{eq:CmaxXab}.
\end{proof}

The new bound is tight and can be saturated exactly. For fixed $ a $ and $ b $, an explicit construction of a state attaining the maximal concurrence is given in ~\ref{app:MaxMix}. 
It is also worth noting that Eq.~\eqref{eq:CmaxXab} provides a more stringent upper bound on the concurrence than that of Refs.~\cite{Zhang:2008ebu,Barr:2024djo}, which gives $\mathcal{C}[\rho] \leq \sqrt{1 - \max\{a^2,b^2\}}$. While this bound is controlled solely by the larger of the two polarization magnitudes $a$ and $b$, Eq.~\eqref{eq:CmaxXab} depends on both simultaneously, and yields a strictly tighter upper bound whenever $a\neq b$ and $\max\{a,b\}<1$.

\subsection{Saturation of the concurrence bound}
\label{rank1} 
We now give an explicit saturation construction for the bound in Eq.~\eqref{eq:CboundAppendix}.
Since $\rho$ must be a mixed state whenever $a \neq b$ (i.e., when the two polarizations differ, or $\Tr \rho_A^2 \neq \Tr \rho_B^2$), it is natural to ask whether the upper bound of the concurrence Eq.~\eqref{eq:CboundAppendix} remains saturable in this setting. We show that the answer is affirmative: a mixed state can indeed reach this bound. From the above derivation, the equality holds for non-pure $\rho$ if all these three conditions are satisfied:
\begin{itemize}
\item $\mathrm{rank}(\mathcal R)=1$ This follows from saturation of the first inequality in Eq.~\eqref{eq:C2bound}, namely, $\mathcal C^2[\rho]=\Tr(\rho\tilde{\rho})=r_1^2$.
\item $\mathrm{rank}(\mathcal \rho)=2$.
\item The eigenvalues of $\rho$ take the boundary values specified in Eq.~\eqref{eq:eigenvaluesOfRho}
\end{itemize}
The second and third conditions are straightforward. Therefore, the crux of this problem lies in determining the condition under which $\mathrm{rank} (\mathcal R) =1 $ for a rank-2 density matrix $\rho$. In \ref{app:MaxMix}, we show that for mixed states, if $\mathrm{rank} (\mathcal R) =1$, up to local unitary transformations, the density matrix $\rho$ takes the form
\begin{equation}
  \rho = \lambda_1 \ket{\psi}\bra{\psi} + \lambda_2\ket{01}\bra{01}
\end{equation}
\begin{equation}
    \ket{\psi}= \alpha \ket{00}+ \beta e^{i\theta} \ket{11}
\end{equation}
\begin{equation}
    \ket{\tilde \psi}= \beta e^{-i\theta} \ket{00} + \alpha \ket{11}
\end{equation}
where $\alpha$ and $\beta$ are positive real numbers with $\alpha^2 + \beta^2 =1 $.
with the corresponding $\mathcal R = \lambda_1^2 \ket{\psi}\langle{\psi}| \tilde \psi\rangle\langle{\tilde \psi}|$. The maximum concurrence is then given by 
$
\mathcal C[\rho] = 2\alpha \beta  \lambda_1
$
In this case, one finds that the maximum concurrence is computed from the product of the first eigenvalue of $\rho$ and the overlap between the corresponding eigenvector and its spin-flipped vector. 

Without loss of generality, let $a>b$. 
One can immediately reproduce the upper bound in Eq.~(\ref{eq:CboundAppendix}) by noting that $\lambda_1 = 1-\frac{a-b}{2}$ and $\lambda_2 = \frac{a-b}{2}$.
Matching the reduced density matrices gives
\begin{equation}
\alpha=\sqrt{\frac{1+b}{2-a+b}},
\qquad
\beta=\sqrt{\frac{1-a}{2-a+b}},
\end{equation}
and the corresponding eigenvectors $|\psi_1\rangle=\frac{1}{\sqrt{2-a+b}}(\sqrt{1+b},0,0,\sqrt{1-a})^T$ and $|\psi_2\rangle=(0,1,0,0)^T$, for the density matrix given in \ref{app:optimal-x}. 

\subsection{Numerical verification and physical interpretation}
\begin{figure}[htbp]
    \centering
    \includegraphics[width=0.7\linewidth]{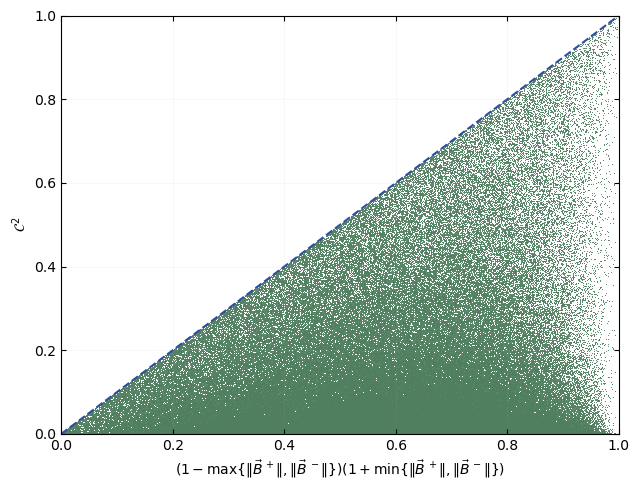}
    \caption{Numerical verification of the boundary of concurrence for general density matrices. The green points are obtained from $2{,}000{,}000$ randomly generated two-qubit density matrices, plotted against $(1-\max\{a,b\})(1+\min\{a,b\})$, while the blue dashed line shows the analytic boundary $\mathcal{C}_{\max}^2=(1-\max\{a,b\})(1+\min\{a,b\})$ in Eq.~\eqref{eq:CmaxXab}.}
    \label{fig:boundary}
\end{figure}

To complement the analytical proof, we perform a numerical
verification of the bound in Eq.~\eqref{eq:CmaxXab} by randomly
generating 2{,}000{,}000 two-qubit density matrices of the form
in Eq.~\eqref{density}. Figure~\ref{fig:boundary} shows the
squared concurrence of these states plotted against the
squared analytical upper bound. All sampled states lie on
or below the boundary, providing a numerical consistency
check of our result.

Eq.~\eqref{eq:CmaxXab} shows that the maximum achievable concurrence is controlled jointly by the two local polarization magnitudes. The factor $1-\max\{a,b\}$ reflects the limiting role of the larger polarization, whereas $1+\min\{a,b\}$ shows that the smaller polarization also contributes to the attainable upper bound. 

\begin{figure}[htbp]
    \centering
    \includegraphics[width=0.7\linewidth]{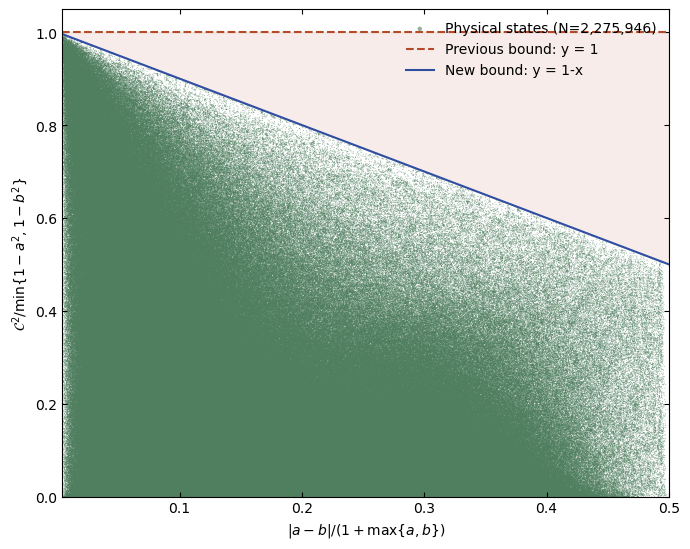}
\caption{Numerical test of the concurrence bound. The figure shows
2{,}275{,}946 two-qubit density matrices with polarizations restricted to
$0.5<a^2<1$ and $0<b^2<0.49$, of which 2{,}092{,}502 are independent random
states and 183{,}444 are generated by random perturbations. The horizontal axis
is $|a-b|/(1+\max\{a,b\})$ and the vertical axis is
$\mathcal{C}^2/\min\{1-a^2,1-b^2\}$. The dashed line shows the bound of
Refs.~\cite{Zhang:2008ebu,Barr:2024djo}, and the solid line shows
Eq.~\eqref{eq:CmaxXab}. No state exceeds the solid line, and the shaded
region between the two lines is never reached.}
    \label{fig:test}
\end{figure}

While Fig.~\ref{fig:boundary} provides a broad numerical
check of our bound, Fig.~\ref{fig:test} highlights its improvement over the previous bound of Refs.~\cite{Zhang:2008ebu,Barr:2024djo}. To display both
bounds in a common representation, we introduce the
normalized coordinates
\begin{equation}
    x=\frac{|a-b|}{1+\max\{a,b\}},
    \qquad
    y=\frac{\mathcal{C}^2[\rho]}{\min\{1-a^2,1-b^2\}}.
\end{equation}
Writing $M=\max\{a,b\}$ and $m=\min\{a,b\}$, we have
$|a-b|=M-m$ and $\min\{1-a^2,1-b^2\}=1-M^2$.
For $M<1$, the bound
$\mathcal{C}[\rho]\leq\sqrt{1-M^2}$
of Refs.~\cite{Zhang:2008ebu,Barr:2024djo} therefore gives
\begin{equation}
    y=\frac{\mathcal{C}^2[\rho]}{1-M^2}\leq 1.
\end{equation}
Meanwhile, Eq.~\eqref{eq:CmaxXab},
$\mathcal{C}^2[\rho]\leq(1-M)(1+m)$, yields
\begin{equation}
    y\leq\frac{(1-M)(1+m)}{1-M^2}
    =\frac{1+m}{1+M}
    =1-\frac{M-m}{1+M}
    =1-x.
\end{equation}
Thus, the dashed line $y=1$ represents the previous bound,
while the solid line $y=1-x$ represents
Eq.~\eqref{eq:CmaxXab}. All numerically generated states lie
below the solid line, and none fall in the shaded region
between the two bounds, which is allowed by the previous
bound but excluded by Eq.~\eqref{eq:CmaxXab}.
In the special case of a pure state, one has $a=b$,
and both bounds coincide.

A useful physical interpretation is that local polarization and nonlocal entanglement draw on complementary spin information. A large local polarization indicates that substantial spin information is stored in an individual subsystem, which constrains the amount of nonlocal correlation that can be shared by the pair. The precise constraint is quantified by Eq.~\eqref{eq:CmaxXab}, where the two polarization magnitudes enter asymmetrically through their maximum and minimum.

In general, the local polarizations satisfy $a \neq b$. Such configurations can be realized, for example, by preparing polarized initial states~\cite{Cheng:2025zaw}, to which the above results directly apply. In the following, we consider a concrete example, $e^+ e^- \to Z^0 \to q\bar{q}$, where $a=b$.

\section{Spin Entanglement in $e^+ e^- \to q \bar{q}$ via $Z$ Exchange}
\begin{figure}[H]
\begin{center}
     \includegraphics[width=0.35\columnwidth]{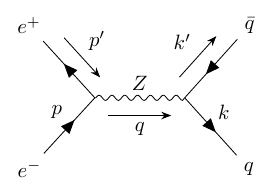}
    \includegraphics[width=0.4\columnwidth]{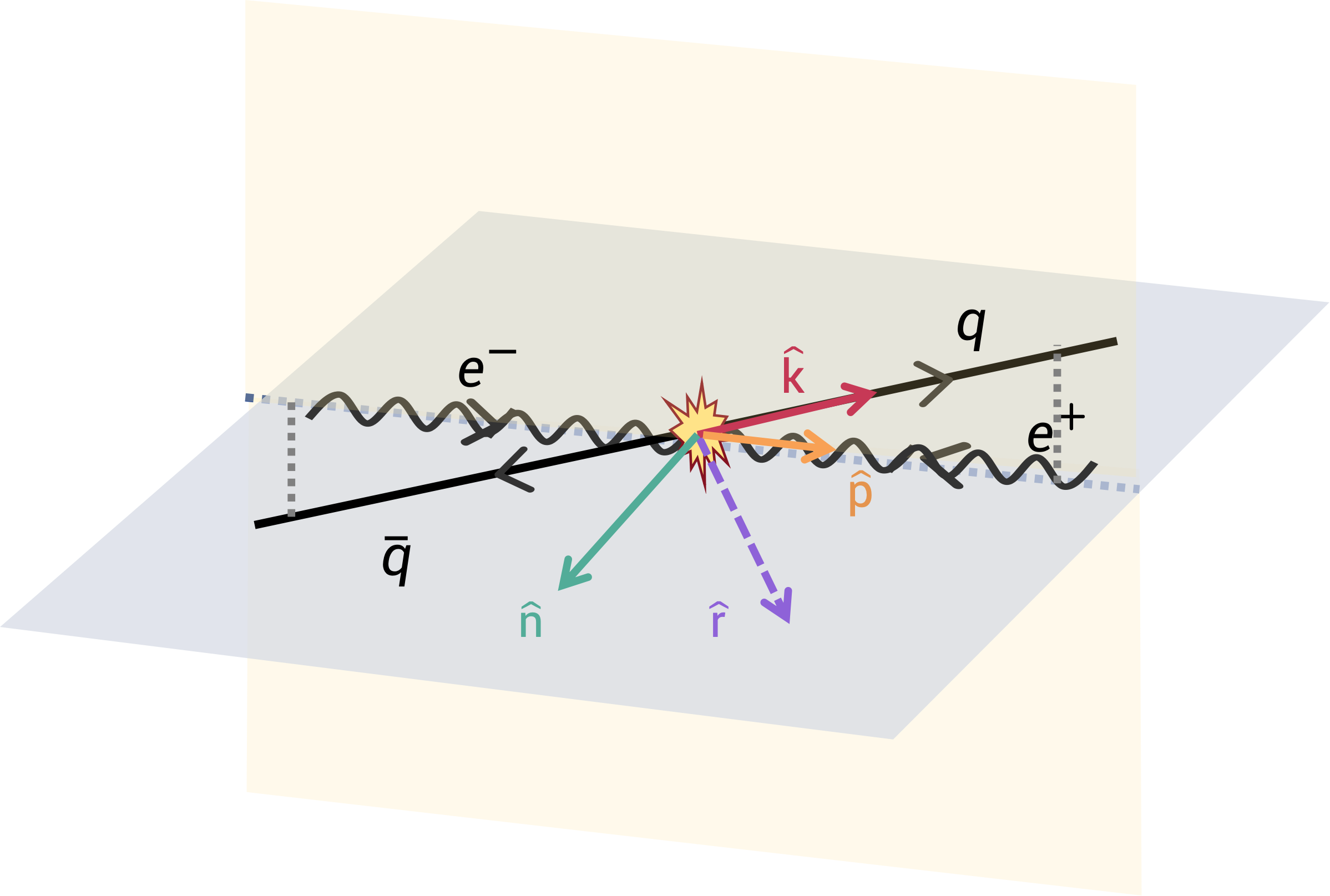}
    \caption{
    (a) Feynman diagram of electron--positron scattering via $Z$ boson.
    (b) Center-of-mass frame for $e^-e^+\to q\bar q$. The outgoing quark direction defines $\hat{\mathbf k}$, the normal to the scattering plane defines $\hat{\mathbf n}$, and $\hat{\mathbf r}$ completes the orthonormal basis $\{\hat{\mathbf n},\hat{\mathbf r},\hat{\mathbf k}\}$.
    } 
    \label{fig:com}
\end{center}  
\end{figure}

We now illustrate the general discussion above with a concrete scattering process.
To introduce polarization, we consider a weak interaction process in which parity is violated. Specifically, we study the process \(e^-(p)\,e^+(p') \;\rightarrow\; q(k)\,\bar q(k'),\) mediated by a $Z$ boson as shown in Fig.~\ref{fig:com}.

Following the procedures of Refs.~\cite{Uwer:2004vp, Afik:2022dgh, Bernreuther:1993hq, Bernreuther:1997gs, Brandenburg:1998xw, Baumgart:2012ay, Bernreuther:2015yna, Afik:2022kwm, Afik:2020onf,  Aoude:2022imd}, we derive the density matrix of the final-state $q\bar q$ pair for a maximally mixed initial state. The unnormalized final-state density matrix, or $R$-matrix, is defined as

\begin{equation}
R_{\alpha \alpha', \beta \beta'} = \frac{1}{4} \sum_{\text{initial spins}}  \mathcal M^\ast_{e^-e^+ \to q^\alpha  \overline{q}^{\alpha^\prime}} \mathcal M_{e^-e^+ \to q^ \beta  \overline{q}^{\beta^\prime} }.
\end{equation}
Writing the $R$-matrix in the basis \{$\alpha \beta,\alpha'\beta'$\}, we obtain the normalized density matrix for the final state as $\rho_f \;=\; \tfrac{R}{\text{Tr} R}$, which takes the general form given in Eq.~\eqref{density}.

The final-state density matrix $\rho_f$ can be evaluated explicitly in the center-of-mass frame using the helicity basis $\{\hat{\mathbf n},\hat{\mathbf r},\hat{\mathbf k}\}$. Here $\hat{\mathbf k}$ denotes the direction of the outgoing quark momentum, $\hat{\mathbf n}$ is perpendicular to the interaction plane, and $\hat{\mathbf r}$ is orthogonal to the $\{\hat{\mathbf n},\hat{\mathbf k}\}$ plane, as illustrated in Fig.~\ref{fig:com}. The Pauli-matrix indices \{1,2,3\} are mapped onto the $\{\hat{\mathbf n},\hat{\mathbf r},\hat{\mathbf k}\}$ coordinates, respectively.

In the center-of-mass frame, let $p^\mu$ and $p'^{\mu}$ be the four-momenta of 
the incoming electron and positron, while $k^\mu$ and $k'^{\mu}$ denote those of the outgoing quark and antiquark. The Mandelstam variable is defined as
\begin{equation}
s = (p+p')^2 = (k+k')^2 ,
\end{equation}
which represents the squared center-of-mass energy. We express all four-momenta in units of $\sqrt{s}$ and define the following two dimensionless variables:
\begin{equation}
u = \sqrt{1 - \frac{4m^2}{s}}, 
\qquad 
z = \hat{\mathbf p}\!\cdot\!\hat{\mathbf k} = \cos\theta ,
\end{equation}
where $u$ represents the speed of the produced quarks in the center-of-mass
frame, and $\theta$ is the angle between $\hat{\mathbf k}$ and $\hat{\mathbf p}$.

\begin{figure}[H]
    \centering    
    \includegraphics[width=0.8\linewidth]{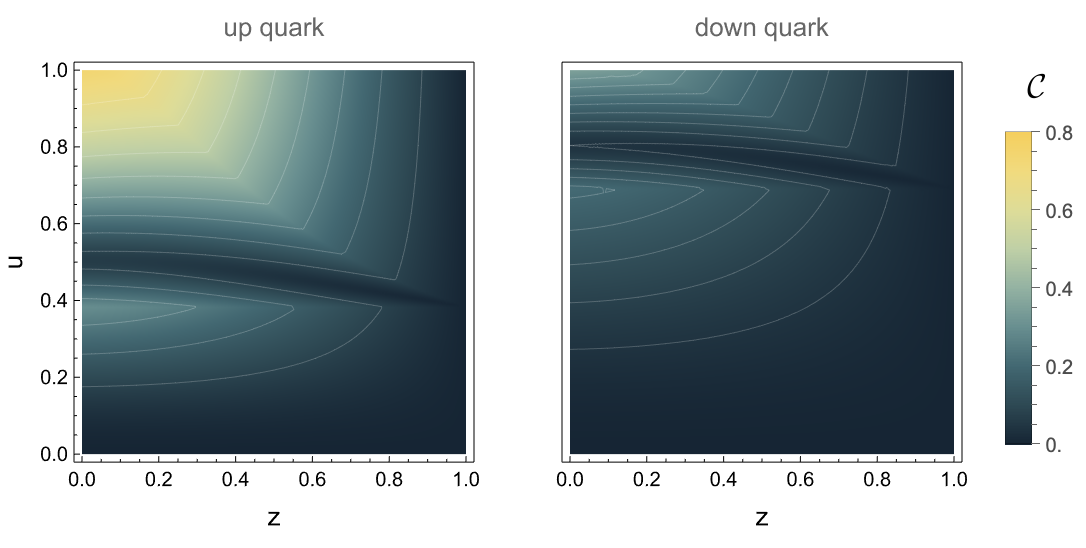}  
    \caption{Concurrence of the final-state $q\bar q$ pair in $e^-e^+$ annihilation near the $Z^0$ pole as a function of the quark speed $u$ and scattering angle variable $z=\cos\theta$. The left and right panels correspond to up-type and down-type quarks, respectively.}
    \label{fig:eeqq}
\end{figure}

From our calculation, the non-vanishing coefficients of the final-state density matrix are
\begin{equation}
\big(C_{ij}\big)=
\begin{pmatrix}
C_{11} & 0      & 0\\[4pt]
0      & C_{22} & C_{23}\\[4pt]
0      & C_{23} & C_{33}
\end{pmatrix},
\qquad
\vec{B}^{\,+} = \vec{B}^{\,-}
=
\begin{pmatrix}
0 \\[4pt]
B_2 \\[4pt]
B_3
\end{pmatrix}.
\end{equation}
The explicit expressions for these coefficients, $B_2$, $B_3$, and $C_{ij}$, are presented in \ref{App. R-matrix}. These results are consistent with those given in Ref.~\cite{Guo:2026yhz}.\footnote{During the preparation of this work, we noted that Ref.~\cite{Guo:2026yhz} appeared with results consistent with what we have obtained. The focus of the present work is nonetheless distinct, as we emphasize the relation between the polarization vectors and the maximal concurrence.} Furthermore, we evaluated the concurrence of final states. The results for up-type and down-type quarks are shown in Fig.~\ref{fig:eeqq}. 

We find that the concurrence reaches its maximum at $u=1$ and $z=0$, given by
\begin{equation}
\mathcal{C}_{\text{max}}
=
\frac{\left(c_A^{q}\right)^2-\left(c_V^{q}\right)^2}
{\left(c_A^{q}\right)^2+\left(c_V^{q}\right)^2},
\label{ce}
\end{equation}
where $c_A^{q}=T_3^{q}$ ($1/2$ for $u,c,t$ quarks and $-1/2$ for $d,s,b$ quarks) and $c_V^{q}=T_3^{q}-2Q_q\sin^2\theta_W$ (approximately $0.1912$ for $u,c,t$ quarks and $-0.3459$ for $d,s,b$ quarks) are the axial-vector and vector couplings of quark $q$ to the $Z$ boson, respectively. Here $T_3^{q}$ is the third component of weak isospin, $Q_q$ is the electric charge of the quark in units of $|e|$, and $\theta_W$ is the Weinberg weak mixing angle ($\sin^2\theta_W=0.2312$). At this point, the explicit form of the final state density matrix \(\rho_f^{opt}\) is shown in \ref{App. R-matrix} which can be verified to be a pure state, and the Bloch vectors of $\rho_f^{opt}$ are:
\begin{equation}
\vec{B}^{\, +}_{opt} = \vec{B}^{\, -}_{opt}
=
\begin{pmatrix}
0 \\[4pt]
0 \\[4pt]
-\dfrac{2c_V^{q}c_A^{q}}
{\left(c_V^{q}\right)^2+\left(c_A^{q}\right)^2}
\end{pmatrix}.
\end{equation}
Since \(\left(c_A^{q}\right)^2 > \left(c_V^{q}\right)^2\), we obtain
\begin{equation}
\sqrt{1-|\vec{B}^{\, \pm}_{opt}|^2}
=
\frac{\left(c_A^{q}\right)^2-\left(c_V^{q}\right)^2}
{\left(c_A^{q}\right)^2+\left(c_V^{q}\right)^2},
\end{equation}
which reproduces Eq.~\eqref{ce}.% and confirms the second case we discussed above.

The maximal values in Fig.~\ref{fig:eeqq} are
\(
\mathcal{C}^{\text{up}}_{\max}\simeq 0.744
\)
and
\(
\mathcal{C}^{\text{down}}_{\max}\simeq 0.353\) which is consistent with Eq.~\eqref{ce} .
Moreover, notice that they are significantly smaller than the maximal concurrence $\mathcal{C}_{\text{max}} =1$ corresponding to the unpolarized case obtained in Ref.~\cite{Qi:2025onf}.

\section{Conclusions and outlook}

In this paper, we establish a quantitative relation between local spin polarization and quantum entanglement in two-qubit systems, and derive an exact analytical expression for the maximal concurrence at fixed polarization magnitudes. In particular, the states that maximize the concurrence are pure states in certain cases, revealing a direct connection between polarization, entanglement, and pure states. Physically, local polarization constrains the amount of nonlocal spin correlation available to the pair, with the analytic bound depending on both polarization magnitudes through $\max\{a,b\}$ and $\min\{a,b\}$.

We illustrate these general results using the parity-violating process $e^+ e^- \to Z^0 \to q\bar{q}$. We find that the concurrence reaches its upper bound at specific kinematic configurations, namely in the ultra-relativistic limit with a perpendicular scattering angle $\theta = \pi/2$. The corresponding maximal concurrence is strictly smaller than unity, with values $\mathcal C_{\max} \simeq 0.744$ for up-type quarks and $\mathcal C_{\max} \simeq 0.353$ for down-type quarks. At these points, the final-state density matrix is a pure state with nonvanishing longitudinal polarization, consistent with the general analysis.

\section*{Acknowledgements}
We thank Zijing Guo, Lingjie Wang, Fanqihang Zhang, Chang Su, Yaoyu Cheng, and Sixuan Gu for useful discussions and comments. This work was supported in part by the Ministry of Science and Technology of China under Grant No. 2024YFA1611004, by the Natural Science Foundation of Guangdong Province under Grant No.~2026A1515011242, and by the CUHK (Shenzhen) University Development Fund under Grant No.~UDF01001859. 
%\end{acknowledgments}

\appendix
\section{Optimal $X$-state density matrices at fixed local polarizations}
\label{app:optimal-x}
This appendix gives the explicit form of optimal density matrices for $X$ states that saturate the bound in Eq.~\eqref{eq:CmaxXab}. When the optimum is reached at $c=1-|a-b|$, one of the two middle diagonal entries vanishes: $\rho_{33}=0$ for $a\ge b$ and $\rho_{22}=0$ for $b\ge a$. In either case, the constraint from Eq.~\eqref{eq:Xstate-positivity} requires $\rho_{23}=0$. The maximal concurrence is then obtained by taking $|\rho_{14}|=\sqrt{\rho_{11}\rho_{44}}$. If $a \ge b$, the optimal density matrix takes the form
\begin{equation}
\rho_X^{\mathrm{opt}}=
\begin{pmatrix}
\dfrac{1+b}{2} & 0 & 0 & \dfrac12\sqrt{(1-a)(1+b)}\\
0 & \dfrac{a-b}{2} & 0 & 0\\
0 & 0 & 0 & 0\\
\dfrac12\sqrt{(1-a)(1+b)} & 0 & 0 & \dfrac{1-a}{2}
\end{pmatrix}.
\end{equation}
For $b \ge a$, one simply swaps $a \leftrightarrow b$ and $\rho_{22} \leftrightarrow \rho_{33}$ in the above expression.

\section{Rank-2 saturation construction for the concurrence bound}
\label{app:MaxMix}

This appendix supports the saturation discussion in Sec.~\ref{rank1} by giving an explicit rank-2 mixed-state construction. We show that imposing \(\mathrm{rank}(\mathcal R)=1\) leads to a family of mixed states that saturate the concurrence bound in Eq.~\eqref{eq:CboundAppendix}.

By the spectral theorem, any rank-2 density matrix can be written as
\begin{equation}
\rho = \lambda_1\ket{\psi_1}\bra{\psi_1}+ \lambda_2 \ket{\psi_2}\bra{\psi_2}
\end{equation}
where $\langle \psi_i | \psi_j \rangle = \delta_{ij}$ and $\lambda_1, \lambda_2 > 0$ as $\rho$ is Hermitian and positive definite.
The spin-flipped density matrix is
\begin{equation}
    \tilde \rho=(\sigma_y \otimes \sigma_y) \rho^* (\sigma_y \otimes \sigma_y) =\Theta \rho \Theta^{-1}
\end{equation}
with $\Theta \equiv -(\sigma_y \otimes \sigma_y) K $ and $K$ is complex conjugation.
Define $\ket{\tilde \psi_i} \equiv \Theta \ket{\psi_i}$, then
\begin{equation}
    \tilde \rho= \lambda_1 \ket{\tilde \psi_1}\bra{\tilde \psi_1} + \lambda_2 \ket{\tilde \psi_2}\bra{\tilde \psi_2}.
\end{equation}
As $\mathrm{rank} (\mathcal R) \le \mathrm{rank}(\rho)=2$, by choosing an appropriate basis, the matrix $\mathcal R =\rho\tilde{\rho}$ effectively reduces to the $2\times2$ matrix:
\begin{equation}
    \begin{pmatrix}
        \lambda_1^2 \langle \psi_1| \tilde \psi_1 \rangle &  \lambda_1 \lambda_2 \langle \psi_1 | \tilde \psi_2 \rangle\\
        \lambda_1 \lambda_2 \langle \psi_2 | \tilde \psi_1\rangle& \lambda_2^2 \langle \psi_2 | \tilde \psi_2 \rangle
    \end{pmatrix} = D \hat{\mathcal R} D
\end{equation} 
where $D = \mathrm{diag}(\lambda_1, \lambda_2)$ and
\begin{equation}
    \hat{\mathcal R} =
    \begin{pmatrix}
        \langle \psi_1| \tilde \psi_1 \rangle &   \langle \psi_1 | \tilde \psi_2 \rangle\\
         \langle \psi_2 | \tilde \psi_1\rangle&  \langle \psi_2 | \tilde \psi_2 \rangle
    \end{pmatrix}.
\end{equation}
Hence $\mathrm{rank}(\mathcal R) = \mathrm{rank}(\hat{\mathcal R})$. In the computational basis
\begin{equation}
    \Theta \ket{00}= \ket{11}, \quad
    \Theta \ket{01}= -\ket{10}, \quad
    \Theta \ket{10}= -\ket{01}, \quad
    \Theta \ket{11}= \ket{00}.
\end{equation}
For convenience, we choose the magic basis
\begin{equation}
\ket{e_1} = \ket{\Phi^+},\quad
\ket{e_2} = i\ket{\Phi^-},\quad
\ket{e_3} = i\ket{\Psi^+},\quad
\ket{e_4}= \ket{\Psi^-}
\end{equation}
with Bell states $\ket{\Phi^\pm}= (\ket{00}\pm\ket{11})/\sqrt{2}$ and $\ket{\Psi^\pm}=(\ket{01}\pm\ket{10})/\sqrt{2}$.
Hence $ \ket{\tilde{e}_i}= \Theta \ket{e_i} = \ket{e_i}$ and $\tilde \rho = \rho^*$ in the magic basis. With local unitary transformations, we can always rotate the basis such that $\ket{\psi_1}$ lies in the plane of two basis vectors.
Without loss of generality, let
\begin{align}
    \ket{\psi_1} &= \alpha^*_1 \ket{e_1}+ \alpha^*_2 \ket{e_2}, \\
    \ket{\tilde \psi_1} &= \alpha_1 \ket{e_1} + \alpha_2 \ket{e_2}.
\end{align}
Since $\langle \psi_1 | \psi_2\rangle = 0$, we can write\footnote{It is tempting to write $\ket{\psi_2}=\alpha_2 \ket{e_1}- \alpha_1 \ket{e_2}$ so that $\langle \psi_1 | \psi_2\rangle = 0$; when $\ket{\psi_2}$ also lies in the $\ket{e_1}$–$\ket{e_2}$ plane, one can show that $\mathrm{rank}(\hat{\mathcal{R}}) = 2$ in this case.}
\begin{align}
       \ket{\psi_2} &= \alpha^*_3 \ket{e_3}+ \alpha^*_4 \ket{e_4}, \\
    \ket{\tilde \psi_2} &= \alpha_3\ket{e_3} + \alpha_4 \ket{e_4}.
\end{align}
Therefore, $\langle \psi_1 | \tilde \psi_1 \rangle = \alpha_1^2+\alpha_2^2$, $\langle \psi_2 | \tilde \psi_2 \rangle= \alpha_3^2+\alpha_4^2$, and $\langle\psi_1|\tilde\psi_2\rangle = 0$. Then $\hat{\mathcal{R}}$ is diagonal
\begin{equation}
\hat{\mathcal R}= 
\begin{pmatrix}
 \langle \psi_1 | \tilde \psi_1 \rangle & 0 \\
 0 & \langle \psi_2 | \tilde \psi_2 \rangle
\end{pmatrix}
\end{equation}

If $\mathrm{rank}(\hat{\mathcal R}) = 1$, either $\langle \psi_1 | \tilde \psi_1 \rangle=0$ or $\langle \psi_2 | \tilde \psi_2 \rangle=0$.  Without loss of generality, choose $\langle \psi_1 | \tilde \psi_1 \rangle\ne0$, $\langle \psi_2 | \tilde \psi_2 \rangle=0$ and $\alpha_4 = i \alpha_3$. We obtain
\begin{equation}
    \ket{\psi_2}= \frac{e^{i\phi}}{\sqrt{2}}(\ket{e_3} + i\ket{e_4})
    = i e^{i\phi}\ket{01}
\end{equation}
As $\alpha_1^2 + \alpha_2^2 \ne 0$, $\ket{\psi_1}$ is a nontrivial linear combination of $\ket{00}$ and $\ket{11}$:
\begin{equation}
    \ket{\psi_1}= \alpha \ket{00}+ \beta e^{i\theta} \ket{11} \equiv\ket{\psi}
\end{equation}
where $\alpha$ and $\beta$ are positive real numbers with $\alpha^2 + \beta^2 =1 $.

Therefore, for mixed states, if $\mathrm{rank} (\mathcal R) =1$, up to local unitary transformations, the density matrix $\rho$ takes the form
\begin{equation}
  \rho = \lambda_1 \ket{\psi}\bra{\psi} + \lambda_2\ket{01}\bra{01}
\end{equation}
with the corresponding $\mathcal R = \lambda_1^2 \ket{\psi}\langle{\psi}| \tilde \psi\rangle\langle{\tilde \psi}|$. The maximum concurrence is then given by 
\begin{equation}
\mathcal C[\rho] = r_1 = \sqrt{\Tr \mathcal R }  =\lambda_1 | \langle\psi|\tilde \psi\rangle| = 2\alpha \beta  \lambda_1
\end{equation}
with $\langle\psi| \tilde \psi\rangle = 2\alpha \beta e^{-i\theta }$. In this case, one finds that the maximum concurrence is computed from the product of the first eigenvalue of $\rho$ and the overlap between the corresponding eigenvector and its spin-flipped vector.

\section{Full expression of final-state density matrix}
\label{App. R-matrix}
This appendix collects the explicit coefficients of the final-state density matrix used in the $e^-e^+\to q\bar q$ example discussed in the main text.

In the helicity basis $\{\hat{\mathbf n},\hat{\mathbf r},\hat{\mathbf k}\}$, the two-qubit spin density matrix is written as
\begin{equation}
\rho_f = \frac{1}{4}\left(
\one_{4}
+  B_i^{+}\, \sigma_i \otimes \one_{2}
+  B_j^{-}\, \one_{2} \otimes \sigma_j
+  C_{ij}\, \sigma_i \otimes \sigma_j
\right).
\end{equation}
For compactness, define the common denominator
\begin{equation}
\mathcal{D} \equiv\;
8\,c_A^e c_V^e c_A^q c_V^q \, z \, u
\nonumber + ({c_A^e}^2 + {c_V^e}^2)
\!\left[(c_A^q)^2 (1+z^2)u^2
+(c_V^q)^2\!\left(2+(-1+z^2)u^2\right)\right],
\end{equation}
The non-vanishing correlation coefficients are
\begin{equation}
\big(C_{ij}\big)=
\begin{pmatrix}
C_{11} & 0      & 0\\[4pt]
0      & C_{22} & C_{23}\\[4pt]
0      & C_{23} & C_{33}
\end{pmatrix},
\end{equation}
with
\begin{align}
C_{11} &=
\frac{\big[(c_A^e)^2+(c_V^e)^2\big]
\big[(c_A^q)^2-(c_V^q)^2\big]
(1-z^2)u^2}{\mathcal{D}},
\\[6pt]
C_{22} &=
-\frac{\big[(c_A^e)^2+(c_V^e)^2\big]
(1-z^2)
\Big[(c_A^q)^2u^2+(c_V^q)^2(-2+u^2)\Big]}{\mathcal{D}},
\\[6pt]
C_{23} &=
\frac{2\,c_V^q
\Big[\big((c_A^e)^2+(c_V^e)^2\big)c_V^q\,z
+2\,c_A^e c_V^e c_A^q\,u\Big]
\sqrt{(1-z^2)(1-u^2)}}{\mathcal{D}},
\\[8pt]
C_{33} &=
\frac{8\,c_A^e c_V^e c_A^q c_V^q \, z \, u
+({c_A^e}^2 + {c_V^e}^2)
\!\left[(c_A^q)^2 (1+z^2)u^2
+(c_V^q)^2\!\left(u^2-z^2(-2+u^2)\right)\right]}{\mathcal{D}}.
\end{align}
The Bloch vectors (polarizations) are identical for the quark and antiquark,
\begin{equation}
\vec{B}^{\, +} = \vec{B}^{\, -}
=
\begin{pmatrix}
0 \\[4pt]
B_2 \\[4pt]
B_3
\end{pmatrix},
\end{equation}
with
\begin{equation}
\begin{aligned}
B_2 &=
-\frac{
2\,c_V^q
\Big[2\,c_A^e c_V^e c_V^q
+\big((c_A^e)^2+(c_V^e)^2\big)c_A^q z u\Big]
\sqrt{(1-z^2)(1-u^2)}
}{\mathcal{D}},
\\[8pt]
B_3 &=
-\frac{
2\Big[
\big((c_A^e)^2+(c_V^e)^2\big)c_A^q c_V^q (1+z^2)u
+2\,c_A^e c_V^e z \big((c_V^q)^2+(c_A^q)^2u^2\big)
\Big]
}{\mathcal{D}} .
\end{aligned}
\end{equation}

At the special point $u=1$ and $z=0$, where the concurrence reaches its maximum in the main text, the density matrix simplifies to
\begin{equation}
\label{app:rho_f}
\rho_f^{opt} =
\begin{pmatrix}
\dfrac{(c_A^q-c_V^q)^2}
{2\left[(c_A^q)^2+(c_V^q)^2\right]}
& 0 & 0 &
\dfrac{(c_A^q-c_V^q)(c_A^q+c_V^q)}
{2\left[(c_A^q)^2+(c_V^q)^2\right]}
\\
0 & 0 & 0 & 0 \\
0 & 0 & 0 & 0 \\
\dfrac{(c_A^q-c_V^q)(c_A^q+c_V^q)}
{2\left[(c_A^q)^2+(c_V^q)^2\right]}
& 0 & 0 &
\dfrac{(c_A^q+c_V^q)^2}
{2\left[(c_A^q)^2+(c_V^q)^2\right]}
\end{pmatrix}.
\end{equation}


%\bibliographystyle{elsarticle-num} 
%\bibliography{refs}


\end{document}